\documentclass[aps,prl,twocolumn,superscriptaddress]{revtex4-2}

\usepackage{graphicx}  
\graphicspath{{img/}}
\usepackage{amsmath}   
\usepackage{amssymb}   
\usepackage{hyperref}  

\usepackage{pdfpages} 
\usepackage{pgffor} 

\makeatletter
\AtBeginDocument{\let\LS@rot\@undefined}
\makeatother

\def\supplementfilename{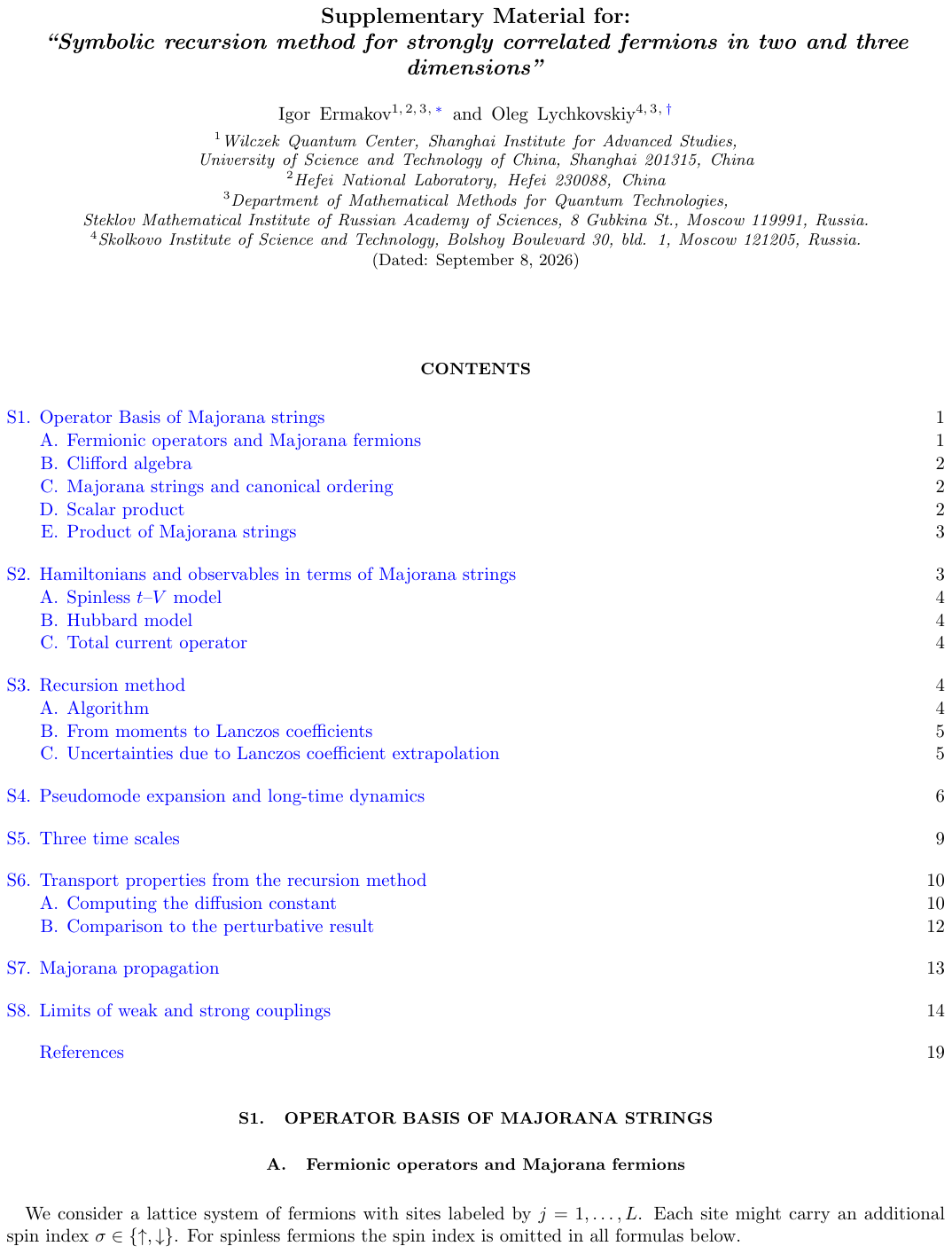}

\pdfximage{\supplementfilename}
\def\numbersupplementpages{\the\pdflastximagepages}

\newif\ifarXiv
\arXivtrue 

\begin{document}

\title{
Symbolic recursion method for strongly correlated fermions \\
in two and three dimensions
}

\author{Igor Ermakov}
\email{ermakov1054@yandex.ru}
\affiliation{Wilczek Quantum Center, Shanghai Institute for Advanced Studies, University of Science and Technology of China, Shanghai 201315, China}
\affiliation{Hefei National Laboratory, Hefei 230088, China}
\affiliation{Department of Mathematical Methods for Quantum Technologies, Steklov Mathematical Institute of Russian Academy of Sciences, 8 Gubkina St., Moscow 119991, Russia.}

\author{Oleg Lychkovskiy}
\email{lychkovskiy@gmail.com}
\affiliation{Skolkovo Institute of Science and Technology, Bolshoy Boulevard 30, bld. 1, Moscow 121205, Russia.}
\affiliation{Department of Mathematical Methods for Quantum Technologies, Steklov Mathematical Institute of Russian Academy of Sciences, 8 Gubkina St., Moscow 119991, Russia.}

\date{\today}

\begin{abstract}
We present a symbolic implementation of the recursion method for dynamical correlations and transport in fermionic systems on one-, two-, and three-dimensional lattices. The implementation is applicable in the strongly correlated regime, yields results directly in the thermodynamic limit, and covers all time scales, from short times through the thermalization time to the late-time asymptotics. Focusing on two paradigmatic models -- interacting spinless fermions and the Hubbard model -- we confirm the universal operator growth hypothesis in the fermionic case, compute infinite-temperature current-current autocorrelation functions, and determine the charge diffusion constant and the high-temperature conductivity. The diffusion constant is obtained in both the perturbative and the nonperturbative regime of the interaction strength, and we locate the boundary between them. We compare our approach with the Majorana propagation method, which converges up to intermediate times and agrees with ours where converged. Our results highlight a symbolic computational paradigm in which the most expensive step is performed once, producing reusable symbolic output that yields physical observables for arbitrary model parameters.

\end{abstract}

\maketitle



Dynamics and transport in strongly correlated electron systems, such as the spinless-fermion $t$-$V$ model and the Hubbard model, have been at the forefront of condensed-matter physics for decades \cite{laughlin1983anomalous,dagotto1994correlated,imada1998metal,tokura2003correlated,troyer2005computational,Brown_2019_Bad,nichols2019spin,guardado2020subdiffusion,Kovacevic_2025_Toward,pham2026spin}. Accurate simulation of these models remains one of the central challenges of modern theoretical physics. Approaches based on the direct diagonalization of Hamiltonians are fundamentally limited by the exponential growth of the many-body Hilbert space with the system size \cite{yamada200516,innerberger2020electron}. In one dimension, remarkable progress has been achieved using 
integrability-based techniques \cite{prosen2011exact,bertini2021finite} and tensor-network methods 
\cite{Wall_2012}, to the point that a broad range of dynamical and transport problems can now be considered simulable in 1D. In higher dimensions substantial progress has been achieved using a variety of analytical and numerical approaches \cite{barthel2009contraction,kraus2010fermionic,corboz2010simulation,pivzorn2010fermionic,bergeron2011optical,Shakirov_2015_Modeling,perepelitsky2016transport,Huang_2019_Strange,ulaga2022thermal,ulaga2023thermoelectric,miller2025simulation,Kovacevic_2025_Toward,d2026majorana}. In parallel, considerable efforts have focused on quantum simulation platforms, which aim to  realize synthetic correlated many-body systems and probe their properties experimentally 
\cite{schreiber2015observation,arute2020observation,vilchez2025extracting,chowdhury2025quantum,alam2025fermionic}. Despite impressive recent progress, such platforms remain at an early stage of development and would greatly benefit from reliable computational benchmarks capable of guiding both experiments and theory.

\begin{figure}[t]
    \centering
    \includegraphics[width=\columnwidth]{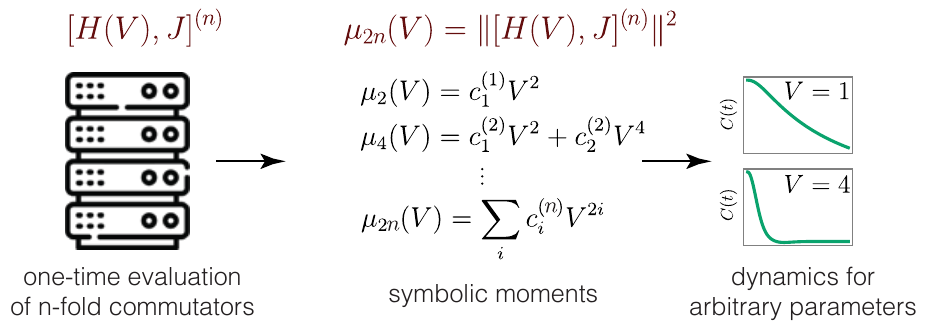}
    \caption{%
Reusable symbolic computation workflow: the computationally expensive symbolic evaluation of the $n$-fold commutators $[H(V),A]^{(n)}$ is carried out once, while the resulting symbolic moments $\mu_{2n}(V)$ are subsequently reused multiple times to compute correlation functions $C(t)$ at specific values of the parameter $V$.
    }
    \label{fig:idea}
\end{figure}

In this work, we develop a symbolic computational approach to the dynamics of strongly correlated fermionic systems based on the recursion method. While first introduced decades ago \cite{Mori_1965_Continued-fraction,Dupuis_1967_Moment,haydock1980recursive,mattis1981reduce,viswanath1994recursion}, the recursion method has seen renewed attention due to recent theoretical progress on operator growth \cite{Elsayed_2014_Signatures,Bouch_2015_Complex,parker2019universal,Avdoshkin_2020_Euclidean,De_2024_Stochastic,Teretenkov_2025_Pseudomode,nandy2025quantum,Gamayun_2025_Exactly,loizeau2025opening,Pinna_2025_Approximation,lunt2025emergent} and advances in computational capabilities.

The recursion method has been successfully applied in a variety of settings, including interacting spin-$1/2$ systems \cite{viswanath1994ordering,Florencio_1992_Quantum,Lindner_2010_Conductivity,yates2020lifetime,yates2020dynamics,Uskov_2024_Quantum,Heveling_2022_Numerically,fullgraf2025lanczos,Loizeau_2025_Quantum,Loizeau_2025_Codebase,Shirokov_2026_Recursion} and higher-spin models \cite{ermakov2025operator}. Other  methods to address quantum  many-body dynamics in Heisenberg representation have also emerged \cite{Rakovszky_2022_Dissipation-assisted,rudolph2023classical,ermakov2024unified,schuster2025polynomial,Begusic_2025_Real-time,angrisani2025simulating,rudolph2025pauli}, typically formulated within qubit operator spaces where the Pauli-string basis provides a natural framework. 

Fermionic models pose additional challenges. In Monte Carlo methods, these challenges manifest themselves as the infamous sign problem. In operator-based approaches, difficulties arise from the anticommutation relations of fermionic operators on spatially separated sites. A dedicated operator-propagation method for fermions, Majorana propagation, has recently been introduced \cite{miller2025simulation,facelli2026fast,alam2025fermionic,shrikhande2025pauli,d2026majorana}, extending the Pauli-propagation algorithms developed for qubits.



Here, we report the successful application of the advanced recursion method to interacting fermionic systems on one-, two-, and three-dimensional lattices. In one dimension, the recursion method is known to be less efficient than state-of-the-art approaches based on exact diagonalization or matrix product states \cite{Yi-Thomas_2024_Comparing, Uskov_2024_Quantum,Shirokov_2026_Recursion}. For this reason, applying the recursion method in one dimension serves the sole purpose of cross-checking its results against those from exact diagonalization. In contrast, dynamics in two and three spatial dimensions is widely believed to be challenging for established numerical methods. We demonstrate that the advanced recursion method is a powerful technique applicable in higher dimensions, capable of capturing the dynamics over the entire time domain, from short times to thermalization, in the strongly correlated regime and without finite-size effects.

We additionally cross-check the recursion method against an independent approach, Majorana propagation \cite{miller2025simulation,facelli2026fast}. We find that the latter is converged only over a limited time interval, within which the two methods agree, thereby confirming our results.


The rest of the manuscript is organized as follows. In the next section, we introduce specific models under study. Then we discuss the autocorrelation function and its moments. After that, we briefly review the recursion method and the universal operator growth hypothesis (UOGH) \cite{parker2019universal}, present Lanczos coefficients for interacting fermionic models and demonstrate that they are compatible with the UOGH. Subsequently, we present infinite-temperature autocorrelation functions obtained from the recursion method supplemented by the UOGH. Then we apply the recursion method to describe transport in the models under study. Specifically, we compute the diffusion constant and the conductivity for a broad range of interaction strengths. In the End Matter, we describe implementation details, moment expansion, the pseudomode expansion and late-time asymptotics of the autocorrelation function, and the Majorana-propagation benchmark. Further details are relegated to the Supplemental Material \cite{supplement}.

\paragraph{Models and observables.}
We consider two fermionic models on hypercubic lattices.  
The first one is the spinless fermion $t$-$V$ model,
\begin{align}\label{HamTV}
    H_{tV} &=
    -\,t_\text{hop} \sum_{\langle i,j\rangle}
    \left( c_i^\dagger c_j + \text{H.c.} \right)
    + V \sum_{\langle i,j\rangle}
    \Bigl(n_i - \tfrac12\Bigr)\Bigl(n_j - \tfrac12\Bigr).
\end{align}
The second one is the Hubbard model,
\begin{align}\nonumber
H_{\mathrm{Hub}} &=
- t_\text{hop} \sum_{\langle i,j\rangle,\sigma}
\left( c_{i\sigma}^\dagger c_{j\sigma} + \text{H.c.} \right)\\
&~~~+ V \sum_i
\left(n_{i\uparrow}-\tfrac12\right)
\left(n_{i\downarrow}-\tfrac12\right).\label{HamHub}
\end{align}
Here $c^\dagger_{i\sigma}$ and $c_{i\sigma}$ are fermionic creation and
annihilation operators  on site $i$, with  $\sigma=\uparrow,\downarrow$ being the spin index. In the spinless $t$-$V$ model the spin index is omitted. Fermionic operators  satisfy the usual anticommutation relations, $\{c_{i\sigma},c_{j\tau}^\dagger\} = \delta_{ij}\,\delta_{\sigma\tau}$.
The parameters $t_\text{hop}$ and $V$ refer to the hopping amplitude and interaction strength, respectively. We set $t_\text{hop}=1$ unless stated otherwise. The symbol $\langle i,j\rangle$ labels pairs of nearest–neighbour sites. We adopt the convention $\hbar=1$ and $k_{\rm B}=1$ throughout the paper. In one dimension both models are integrable \cite{Lieb_1968_Absence,Takahashi_1999_Thermodynamics}.

We choose the total current along a fixed lattice direction,  $\hat{x}$, as an observable of interest. For the Hubbard model the current reads
\begin{align}\label{totCur}
    J =
    -\, i
    \sum_{i,\sigma}
    \left( c_{i\sigma}^\dagger c_{i+\hat{x},\sigma}
         - c_{i+\hat{x},\sigma}^\dagger c_{i\sigma} \right),
\end{align}
where $i+\hat{x}$ denotes the nearest neighbor of site $i$ in the direction $\hat{x}$ (in one dimension this reduces to $i+1$). For the $t$-$V$ model the current is given by the same formula with subscripts $\sigma$ dropped.

\begin{figure*}[t]
    \centering
    \includegraphics[width=\textwidth]{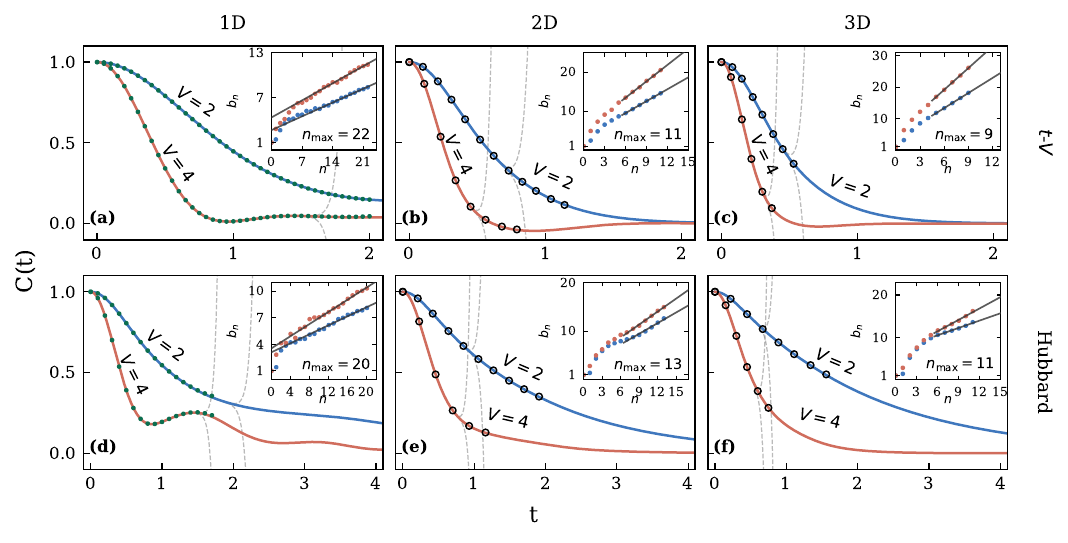}
    \caption{Dynamical infinite-temperature autocorrelation function  (\ref{infC}). Top row: spinless-fermion $t$-$V$ model (\ref{HamTV}). Bottom row: Hubbard model (\ref{HamHub}). Solid lines: results of the recursion method in the thermodynamic limit for the two values of interaction strength $V$. Gray dashed lines indicate upper and lower moment expansion bounds (\ref{bounds}). Green dots in panels (a) and (d)  display the exact diagonalization results for finite periodic chains (of lengths $12$ and $7$, respectively) evaluated for times small enough to be representative for the thermodynamic limit. Black circles show results obtained by Majorana propagation up to the times for which the method remains converged.}
    \label{fig:corrF}
\end{figure*}

\paragraph{Autocorrelation function and moments.} We introduce an inner product in the space of operators according to $(A|B)\equiv \mathrm{tr}(A^\dagger B)/\mathrm{dim}(\mathcal{H})$, where $A$ and $B$ are arbitrary operators and $\mathrm{dim}(\mathcal{H})$ is the (finite) Hilbert space dimension. The corresponding operator norm reads $\lVert A\rVert=\sqrt{(A|A)}$. 

We wish to compute the dynamical infinite-temperature autocorrelation function of the current defined as
\begin{align}\label{infC}
    C(t)\equiv \frac{(J(t)|J)}{\lVert J\rVert^2}.
\end{align}
Here $J(t)=e^{it\mathcal{L}}J$ is the current operator (\ref{totCur}) in the Heisenberg representation, with $\mathcal{L}\equiv[H,\cdot]$ being the Liouvillian superoperator acting on the space of operators. 

Note that since the models we consider conserve the total number of fermions, in the Hubbard model separately for each polarization, the equilibrium state of the system is characterized by average density of fermions, in addition to temperature. The above definition of the autocorrelation function corresponds to half-filling, i.e. to $\langle n_{j}\rangle=1/2$ and $\langle n_{j\sigma}\rangle=1/2$ for the $t$-$V$ and Hubbard models, respectively.


Moments of the autocorrelation function are defined as
\begin{align}\label{moments}
\mu_{2n}\equiv
\frac{(\mathcal{L}^{2n}J|J)}{\lVert J\rVert^2}
=
\frac{(\mathcal{L}^{n}J|\mathcal{L}^{n}J)}{\lVert J\rVert^2}
=(-1)^n \frac {d^{\mkern1mu2n}}{dt^{2n}} C(t)\big|_{t=0}.
\end{align}
Note that this definition implies $\mu_0=1$.

The moments are polynomials in the Hamiltonian parameters. Specifically, for the Hamiltonians~(\ref{HamTV}) and~(\ref{HamHub}), the moments~(\ref{moments}) are polynomials in $V$,  $\mu_{2n}(V)=\sum_{k=1}^{n} c^{(n)}_k V^{2k}$. In case of $t_\text{hop}\neq1$, they scale as $\mu_{2n}(t_\text{hop},V)=(t_\text{hop})^{2n}\mu_{2n}(V/t_\text{hop})$.

Our code  keeps the interaction strength as a symbol, so the first $n_{\text{max}}$ moments come out as exact polynomials in $V$. The computation is performed directly in the thermodynamic limit. We have obtained the coefficients $c^{(n)}_k$ for the Hamiltonians~(\ref{HamTV}) and~(\ref{HamHub}) in one, two and three dimensions. These results are publicly available at \cite{ErmakovSymbolicMoments}. 
%

One can extract certain information on the autocorrelation function $C(t)$ directly from $n_{\text{max}}$ explicitly computed moments. In particular, they yield upper and lower bounds on $C(t)$ \cite{Platz_1973_Rigorous,Roldan_1986_Dynamic,Brandt_1986_High,Bohm_1992_Dynamic,Uskov_2024_Quantum} (see  Fig.~\ref{fig:corrF}), as discussed in the End Matter. These bounds are extremely tight at short times but then abruptly become loose and essentially uninformative.


\paragraph{Lanczos coefficients and UOGH.} The central construction of the recursion method is the Lanczos basis $\{O^0,O^1,O^2,\dots\}$ in operator space, in which the Liouvillian $\mathcal{L}$ is tridiagonal. This basis is constructed recursively as follows: $O^0=J/\|J\|$, $A^1={\cal L} O^0$, $b_1=\|A^1\|$, $O^1=b_1^{-1}\,A^1$ and
$A^n={\cal L}\, O^{n-1}- \,b_{n-1} O^{n-2}$, $b_n=\|A^n\|$, $O^n=b_n^{-1} A^n$ for $n\geq2$, with $A^n$ being unnormalized counterparts of $O^n$. The coefficients $b_n$ are referred to as {\it Lanczos coefficients}. One can straightforwardly check that ${\cal L}$ is indeed tridiagonal in this basis: $(O^m|{\cal L} O^n)=\delta_{m\,n-1} b_n+\delta_{m\,n+1} b_m$.

In fact, Lanczos coefficients $b_n$ can be obtained directly from moments $\mu_{2n}$ \cite{Dupuis_1967_Moment}. We employ this relation reviewed in \cite{supplement} to symbolically compute Lanczos coefficients as functions of the interaction strength for the $t$-$V$ and Hubbard models in one, two and three dimensions, with the results illustrated in Fig.~\ref{fig:corrF}. 

An important question is the asymptotic behavior of Lanczos coefficients at large $n$. It can be proven for locally interacting systems that $b_n$ cannot grow with $n$ faster than linear (with an additional logarithmic correction for one-dimensional systems) \cite{parker2019universal}. The UOGH \cite{parker2019universal} asserts that for chaotic quantum many-body systems  this bound is saturated, i.e.,
\begin{equation}\label{UOGH}
b_n=\alpha n+\gamma+o(1)\qquad {\rm as}\qquad n\rightarrow\infty
\end{equation}
in two and three spatial dimensions.  For integrable systems the scaling is not universal: while typically it is either $\sqrt {n}$ or $O(1)$ \cite{parker2019universal}, linear scaling has also been observed \cite{Dymarsky_2021_Krylov,Bhattacharjee_2022_Krylov,Camargo_2023_Krylov,Avdoshkin_2024_Krylov,
Vasli_2024_Krylov,he2025krylov}.

The UOGH has been previously probed in various spin-1/2 systems on one- \cite{parker2019universal,Noh_2021,Heveling_2022_Numerically,De_2024_Stochastic}, two- \cite{Heveling_2022_Numerically,Uskov_2024_Quantum,De_2024_Stochastic} and three-dimensional \cite{Shirokov_2026_Recursion} lattices, for higher-spin systems on one- and two-dimensional lattices \cite{ermakov2025operator}, as well as for a three-site system of interacting bosons \cite{Bhattacharyya_2023_Operator}. A rigorous proof of the UOGH has been obtained  for a class of  spin-1/2 systems on hypercubic lattices \cite{Cao_2021_Statistical}.  Finally, the UOGH has been confirmed \cite{parker2019universal} for the exactly solvable Sachdev–Ye–Kitaev model \cite{Sachdev_1993_GaplessSpinFluid,Maldacena_2016_RemarksSYK} that describes fermions with random infinite-range couplings. 

Here we probe the  UOGH for locally interacting nonintegrable fermionic systems in two and three dimensions. Our results are clearly consistent with the linear asymptotic scaling of $b_n$ for the 2D and 3D $t$-$V$ model, see the insets in Fig. \ref{fig:corrF} (b), (c), as well as for the 2D Hubbard model, see the inset in Fig. \ref{fig:corrF} (e). 

For the 3D Hubbard model, the linear asymptotics of the Lanczos coefficients is not apparent from our data shown in Fig. \ref{fig:corrF}(f), likely due to the limited number of coefficients we can compute. We therefore treat the asymptotic linear growth of $b_n$ in this model as a working assumption, which we adopt for the subsequent analysis.

Remarkably, we also observe approximately linear scaling for the one-dimensional $t$-$V$ and Hubbard models, which is somewhat unexpected for Bethe-ansatz-integrable models. We leave the discussion of the origin and implications of this fact for further work.

\paragraph{Applying the recursion method.} The recursion method strengthened by the UOGH has proved highly useful in accessing dynamics at times beyond the convergence radius of the Taylor expansion \cite{parker2019universal,Uskov_2024_Quantum,Teretenkov_2025_Pseudomode,Shirokov_2026_Recursion}.  At its core, the method is the solution of the coupled Heisenberg equations in the Lanczos basis. When specified for finding the autocorrelation function, the method amounts to solving coupled differential equations,
\begin{align}\label{diffEqC} 
&\partial_t \varphi_n(t)
= - b_{n+1}\varphi_{n+1}(t)
    + b_n \varphi_{n-1}(t),
\qquad n=0,1,2,\dots 
\end{align}
with the convention  $\varphi_{-1}(t)=0$ and initial condition $\varphi_n(0)=\delta_{n\,0}$. The autocorrelation function $C(t)$ is given by  $\varphi_0(t)$.

However, the number $n_{\rm max}$ of Lanczos coefficients that can be explicitly computed is necessarily limited, since the size of nested commutators ${\cal L}^nJ\equiv[H,J]^{(n)}$ grows exponentially with $n$. If we were to truncate the infinite system of equations at $n=n_{\rm max}$, the resulting approximation would be too rough to  access $C(t)$ up to the thermalization time \cite{supplement}. Fortunately, this limitation can be overcome by  extrapolating the unknown Lanczos coefficients for $n>n_{\rm max}$ according to the UOGH \eqref{UOGH}. This approach has already proven successful for strongly correlated spin  systems \cite{parker2019universal,Uskov_2024_Quantum,Teretenkov_2025_Pseudomode,Shirokov_2026_Recursion}. Here we apply it to fermionic systems. Details of the extrapolation procedure and estimates of its uncertainties are provided in the Supplemental Material \cite{supplement}.

Specifically, we employ the linear fit \eqref{UOGH} to extrapolate the Lanczos coefficients for $n>n_{\text{max}}$, while for $n\leq n_{\text{max}}$ the exact values of $b_n$ are used. The resulting hybrid set of coefficients is then substituted into eq.~(\ref{diffEqC}). The system~(\ref{diffEqC}) is truncated at a finite $K\gg n_{\text{max}}$ and then  solved numerically, yielding the autocorrelation function $C(t)=\varphi_0(t)$. The value of $K$ is chosen such that the dynamics within the considered time window is unaffected by the truncation. For all calculations reported here we use $K=15000$.  

In Fig.~\ref{fig:corrF}, we show the dynamical current-current autocorrelation function~(\ref{infC}) for the models~(\ref{HamTV}) and~(\ref{HamHub}) in one, two, and three spatial dimensions. Results are presented for two representative interaction strengths in the intermediate-to-strong coupling regime, $V=2$ and $V=4$.

We find that, in two and three dimensions, the recursion method yields the autocorrelation function up to thermalization times. The associated uncertainties are analyzed in detail in the Supplemental Material \cite{supplement}. Moreover, supplementing the recursion method with the pseudomode expansion provides access to the leading late-time asymptotics of the correlation function, as discussed in the End Matter and the Supplemental Material.

At short times, the recursion-method results are underpinned by the moment-expansion bounds described in the End Matter. In one dimension, the results are benchmarked against exact diagonalization.

In two and three dimensions, we additionally compare the recursion-method results with an independent Majorana-propagation calculation \cite{miller2025simulation,facelli2026fast}, which does not rely on the UOGH. The details of the calculations are presented in the End Matter and Supplemental Material. 
We find that Majorana propagation converges only up to intermediate times, typically shorter than the thermalization timescale. Wherever the Majorana-propagation results are converged, the two methods agree.


\begin{figure}[t]
    \centering
    \includegraphics[width=\columnwidth]{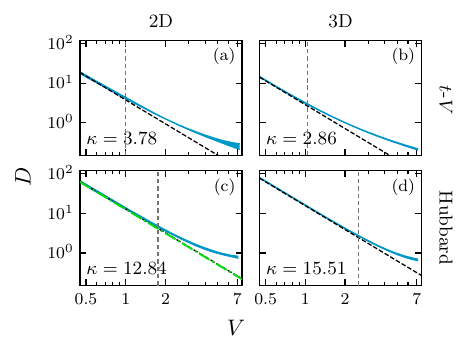}
    \caption{%
    Diffusion constant at half filling and infinite temperature as a function of the interaction strength. Blue band: the result of the recursion method; its width shows the estimated uncertainty. Black dashed line: the fit $\kappa/V^2$
    to the small-$V$ data, with the fitted $\kappa$ quoted in each panel. Green  dashed line in panel (c): the perturbative result of Ref.~\cite{Kovacevic_2025_Toward}; it is accurately captured by our small-$V$ fit. Gray vertical dashed line: the interaction strength at which the diffusion constant deviates from the fit by $10\%$, ending the perturbative $1/V^2$ regime.
    }
    \label{fig:diffusion constant}
\end{figure}

\paragraph{Transport.} The knowledge of current-current autocorrelation function unlocks transport properties of the system through the Kubo linear response theory~\cite{Mahan_2013_Many}. In particular, the direct-current (dc) conductivity, $\sigma_{\rm dc}$, and the charge diffusion constant, $D$, read $\sigma_{\rm dc} =\beta \, {\cal V}^{-1} \, {\rm Re}\int_0^\infty dt \langle J(t) J\rangle_\beta$ and $D =\delta N_{\rm tot}^{-2}\,{\rm Re}\int_0^\infty dt \langle J(t) J\rangle_\beta$ respectively. Here $\beta=1/T$ is the inverse temperature, ${\cal V}$ is the number of lattice sites, $\langle \dots \rangle_\beta$ is the thermal average,  $N_{\rm tot} =\sum_{i,\sigma} n_{i\sigma}$ is the total number of fermions and $\delta N_{\rm tot}^2 = \langle N_{\rm tot}^2 \rangle_\beta-\langle N_{\rm tot} \rangle_\beta^2$. At infinite temperature, $\langle J(t) J\rangle_\beta=\lVert J\rVert^2 C(t)$ with $C(t)$ defined by eq.~\eqref{infC}. The integral over the current-current autocorrelation function can be extracted directly from the Lanczos coefficients \cite{Joslin_1986_Calculation,Wang_2024_Diffusion,Bartsch_2024_Estimation}, which facilitates transport calculations \cite{Wang_2024_Diffusion,Bartsch_2024_Estimation,Uskov_2024_Quantum}. In the high-temperature limit, the diffusion constant saturates at a finite value, while the dc conductivity scales linearly with the inverse temperature, 
\begin{equation}\label{conductivity}
\sigma_{\rm dc} \simeq \beta \, {\cal V}^{-1}\,\left(\left.\delta N_{\rm tot}^2\,D\right|_{\beta=0}\right).
\end{equation}

In Fig.~\ref{fig:diffusion constant}, we present the infinite-temperature diffusion constants for the two- and three-dimensional $t$-$V$ and Hubbard models as functions of the interaction strength $V$ (see Ref.~\cite{supplement} for details of the calculations). Perturbation theory predicts $D = \kappa /V^2$ at small $V$ with some $\kappa>0$ \cite{vuvcivcevic2019conductivity,Kiely_2021_Transport,Vucicevic_2023_Charge,Kovacevic_2025_Toward}. We recover this scaling up to interaction strengths $V\sim 1$ and observe clear deviations at larger $V$.  Our results thus cover both the perturbative and nonperturbative regimes and identify the interaction scale at which perturbation theory ceases to apply.

There exists compelling experimental \cite{Brown_2019_Bad} and theoretical \cite{Brown_2019_Bad,Mousatov_2019_bad,Vranic_2020_Charge,Kiely_2021_Transport,Vucicevic_2023_Charge} evidence that the linear scaling of conductivity with $\beta$, as expressed in Eq. \eqref{conductivity}, remains valid down to relatively low temperatures on the order of $t_{\rm hop}$ or lower. Combined with this observation, our calculations yield results for the dc conductivity across a wide range of interaction strengths and temperatures.

\paragraph{Summary and outlook.}

We have implemented the symbolic recursion method for strongly correlated fermions on two- and three-dimensional lattices, focusing on the $t$-$V$ and Hubbard models. Our implementation yields the moments and Lanczos coefficients of dynamical autocorrelation functions as symbolic expressions in the model parameters. We have verified that the Lanczos coefficients of these locally interacting fermionic models are generally consistent with the universal operator growth hypothesis (UOGH). By supplementing the recursion method with the UOGH, we have computed dynamical correlation functions up to thermalization times. The additional use of the pseudomode expansion has yielded their leading late-time asymptotics. We have also computed transport properties, specifically the infinite-temperature diffusion constant and the dc conductivity at temperatures above the model bandwidth.

We have independently validated our results at short and intermediate times using Majorana propagation, which does not rely on the UOGH. The two methods agree wherever the Majorana-propagation results are converged. Majorana propagation is, however, far more expensive: a single Trotter step can cost as much as our entire symbolic calculation, and reaching $t=2$ for the two-dimensional $t$-$V$ model at $V=2$ required about one month on a 128-core machine with $1$~TB of memory. It is also not symbolic and must be repeated in full for each interaction strength, and even then it converges only up to intermediate times, typically below the thermalization timescale. Our implementation of the recursion method instead reaches all timescales, and its demanding part, the computation of the Lanczos coefficients, is performed symbolically only once, after which correlation functions and transport coefficients follow for any interaction strength within minutes on a laptop.

Our work paves the way for further development of operator-based methods for fermionic systems. An immediate extension would be to reduce the computational complexity by exploiting symmetries beyond the translational symmetry employed here \cite{teng2025leveraging}. It would also be interesting to apply the method to quench dynamics, although recent work has identified obstacles that make the recursion method less efficient for quench dynamics than for correlation functions \cite{Shirokov_2026_Recursion}.

Perhaps the most important direction is to extend the recursion method to low temperatures. Ref.~\cite{Angelinos_2026_Temperature} recently proposed a route to arbitrary temperatures based on nonlinear Toda equations, with the infinite-temperature Lanczos coefficients serving as initial data. This sophisticated and computationally demanding approach has so far been successfully applied only to a finite one-dimensional spin-$1/2$ chain \cite{Angelinos_2026_Temperature}. Extending it to higher-dimensional fermionic systems first requires the computation of sufficiently many infinite-temperature Lanczos coefficients for such systems, precisely the central computational challenge addressed in our work. Our results therefore provide an essential foundation for future applications of the recursion method at low temperatures.

More broadly, this work demonstrates the value of the symbolic computational paradigm, which we expect to be productive well beyond the models considered here.


\begin{acknowledgments}
\paragraph{Acknowledgments.}
The authors are grateful to I. Shirokov for useful discussions. 
\end{acknowledgments}

\paragraph{Data availability.} The core data required to reproduce our results, the symbolic moments \eqref{moments}, are publicly available \cite{ErmakovSymbolicMoments}.

\section{End Matter}

\paragraph{Majorana strings.} Here, we discuss technical aspects of computing commutators in the basis of Majorana strings and the moments~(\ref{moments}). To compute the $n$-fold nested commutator $[H(V),J]^{(n)}$, we employ an operator basis of Majorana strings. A Majorana string is a product of Majorana operators in the canonical order, $\Gamma=\gamma_{i_1,\sigma_1}\gamma_{i_2,\sigma_2}\dots\gamma_{i_l,\sigma_l}$, where $\gamma_{2j-1,\sigma}=c_{j\sigma}+c_{j\sigma}^\dagger$ and $\gamma_{2j,\sigma}=-i(c_{j\sigma}-c_{j\sigma}^\dagger)$. The canonical order implies that (i) the Majorana indices are non-decreasing, $i_1\leq i_2\leq\dots\leq i_l$, and (ii) two operators carrying the same index are ordered with the spin-up one first, i.e., $\gamma_{i,\uparrow}\gamma_{i,\downarrow}$. Since $\gamma_{i,\sigma}^2=1$, no pair $(i,\sigma)$ occurs twice, so these rules order the string completely. For a system of spinless (spinful) fermions  on $\cal V$  sites, there are in total ${\cal D}=4^{\cal V}$ ( ${\cal D}=4^{2{\cal V}}$) canonical
Majorana strings $\Gamma_s$, which form an operator basis
$\mathcal{G}=\{\Gamma_s\}_{s=1}^{{\cal D}}$ orthogonal with respect to the inner product introduced earlier.  

Any fermionic operator can be uniquely decomposed in this basis of Majorana strings. Consequently, the $n$-fold commutator can be written as
\begin{align}\label{operatorDec}
[H(V),J]^{(n)}=\sum_{k=1}^{\mathcal{N}_n} a_k \Gamma_k,
\end{align}
where $a_k$ are complex coefficients and $\mathcal{N}_n$ denotes the number of nonzero Majorana strings appearing at order $n$.

Our routines operate directly in the space of Majorana strings, each stored as an ordered list of indices. We first express the Hamiltonians~(\ref{HamTV},\ref{HamHub}) and the observable~(\ref{totCur}) in this basis, and then iteratively construct the commutators $[H,J]^{(n)}$ until memory constraints are reached. Handling Majorana strings is closely analogous to handling Pauli strings for qubits, with the additional step of restoring the canonical ordering, which costs time but no extra memory.

The coefficients $a_k$ are kept in symbolic form throughout the computation. As a result, the moments $\mu_{2n}(V)$ are obtained as polynomials in $V$, while the Lanczos coefficients are square roots of rational functions of $V$. This extends the symbolic computational approach, previously applied to spin systems \cite{rudolph2023classical,Uskov_2024_Quantum,Teretenkov_2025_Pseudomode,Shirokov_2026_Recursion,Fontana_2025_Classical,rudolph2025pauli,lychkovskiy2026qcommute,PauliStrings,loizeau2026krylov}, to fermionic systems.

\paragraph{Bounds from the moments.} The moments can be used to constrain $C(t)$ from above and from below. The simplest such bound reads \cite{Platz_1973_Rigorous,Roldan_1986_Dynamic,Brandt_1986_High,Bohm_1992_Dynamic,Uskov_2024_Quantum}
\begin{align}\label{bounds}
    P_{4l\pm 2}(t)\leq C(t)\leq P_{4l}(t),
\end{align}
where $P_{2n}(t)=\sum_{m=0}^{n}\big((-1)^m \mu_{2m}/(2m)!\big)\, t^{2m}$ are Taylor polynomials of $C(t)$. These bounds are extremely tight initially but rapidly diverge at later times, and their scope is ultimately limited by the convergence radius $t^\ast$ of the Taylor expansion, which is finite for 2D and 3D systems \cite{parker2019universal}. They are not the central result of the present work, but they serve as a benchmark for the early-time dynamics of $C(t)$, in some cases extending to relatively long times. We provide the moments required to construct them in symbolic form \cite{ErmakovSymbolicMoments}, so that they are readily available for benchmarking other numerical methods and quantum simulations of the $t$-$V$ and Hubbard models at arbitrary interaction strength.

\paragraph{Pseudomode expansion and long-time dynamics.}
A broad class of correlation functions, including current-current autocorrelation functions,  can be represented as a sum over {\it pseudomodes} \cite{garraway1996cavity,Prosen_2002_Ruelle,Fine_2004_Long-time,Morgan_2008_Universal,Sorte_2011_Long-time,Meier_2012_Eigenmodes,Barocchi_2014_Exponential,Mori_2024_Liouvillian-gap,Znidaric_2024_Momentum-dependent,Teretenkov_2025_Pseudomode,Dodelson_2024_Ringdown,Li_2024_Emergent,yoshimura2025theory,Jacoby_2025_Spectral,ozaki2026exact},
\begin{equation}
C(t)=\sum_l (U_l)^2 e^{\lambda_l t},
\label{eq:pseudomode_expansion_main}
\end{equation}
where the generally complex pseudomode eigenvalues $\lambda_l$ encode decay rates and oscillation frequencies, while $U_l$ are the corresponding pseudomode amplitudes. Complex eigenvalues and amplitudes occur in complex-conjugate pairs, ensuring that $C(t)$ is real. The pseudomode (or the complex-conjugate pair of pseudomodes) with the smallest $|{\rm Re}\lambda_l|$ is the {\it leading} pseudomode which determines the asymptotic late-time behavior of the correlation function. Often, truncating the pseudomode expansion to a small number of terms provides a good approximation to the correlation function at all times \cite{Teretenkov_2025_Pseudomode}.

The lower part of the pseudomode spectrum and the corresponding amplitudes can be extracted directly from the Lanczos coefficients within the recursion method \cite{Teretenkov_2025_Pseudomode}. We apply this procedure to the extrapolated Lanczos sequence. Its implementation and the analysis of the associated uncertainties are described in the Supplemental Material \cite{supplement}. Figure~\ref{fig:pseudomodesEM} illustrates the  pseudomode spectrum, as well as the autocorrelation functions obtained from eq. \eqref{eq:pseudomode_expansion_main},  for the two-dimensional $t$-$V$ model.

\begin{figure}[t]
    \centering
    \includegraphics[width=\columnwidth]{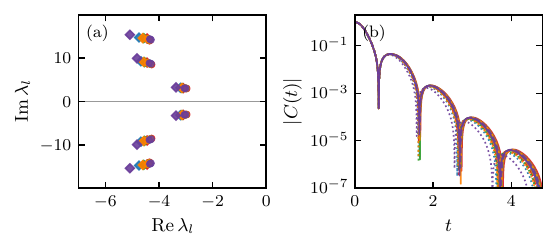}
    \caption{(a) Lower part of the pseudomode spectrum for the two-dimensional $t$-$V$ model at $V=4$. Different markers correspond to variations in the extrapolation of the Lanczos coefficients (see the Supplemental Material \cite{supplement} for details). The  clustering of the eigenvalues demonstrates the robustness of the extracted pseudomodes. (b) The autocorrelation function reconstructed from the truncated pseudomode expansion \eqref{eq:pseudomode_expansion_main}. Colors and line styles label the same variations as in (a).}
    \label{fig:pseudomodesEM}
\end{figure}

\paragraph{Majorana propagation.} To independently check the correlation functions produced by the recursion method, we recompute $C(t)$ with Majorana propagation (MP) \cite{miller2025simulation,facelli2026fast}. MP evolves the current in the Heisenberg picture as an explicit sum of Majorana strings: the Trotterized evolution $e^{iH\,dt}$ is applied as a product of elementary two-Majorana rotations, each of which either leaves a string unchanged or splits it into two. The number of strings grows rapidly under this branching, so the operator is truncated at every step in three ways: in string weight, at a maximum $w$, in coefficient magnitude, at a threshold taken relative to the instantaneous norm of the propagated operator, and in the total number of retained strings, at the allowed operator evolution dimension $\mathcal{N}_{\max}$. MP does not use the UOGH, so the two methods share no common approximation. The algorithm and its parameters are exposed in the Supplemental Material \cite{supplement}.

\begin{figure}[t]
    \centering
    \includegraphics[width=\columnwidth]{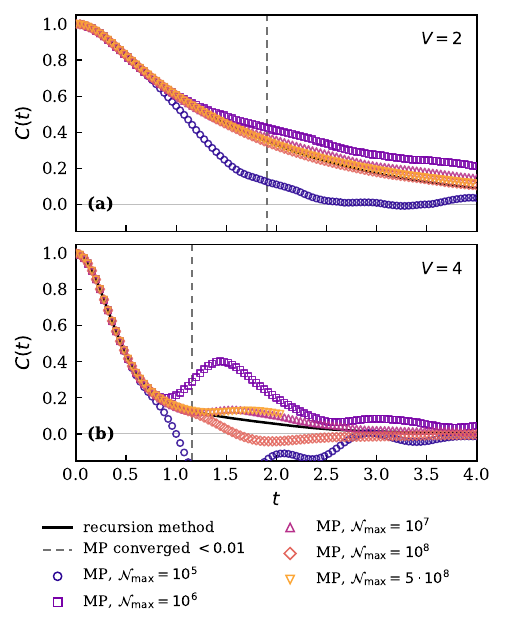}
    \caption{Recursion method versus Majorana propagation (MP) for the infinite-temperature current autocorrelation function $C(t)$ of the 2D Hubbard model at (a) $V=2$ and (b) $V=4$. MP is performed on a finite $16\times16$ lattice with periodic boundary conditions. Black line: recursion method. Markers: MP at allowed operator evolution dimensions $\mathcal{N}_{\max}$ from $10^{5}$ to $5\cdot10^{8}$, all with weight cutoff $w=12$, relative coefficient cutoff $\varepsilon_{\rm rel}=10^{-8}$ and time step $dt=0.01$. Gray dashed vertical line: MP convergence horizon, see text.}
    \label{fig:MP_benchmark}
\end{figure}

Figure~\ref{fig:MP_benchmark} compares the two methods for the 2D Hubbard model at $V=2$ and $V=4$, with $\mathcal{N}_{\max}$ spanning four orders of magnitude. At early times all truncations agree with each other and with the recursion method. Beyond a certain time, each curve departs from the previous one and the recursion result, and the departure point moves to later times as $\mathcal{N}_{\max}$ grows. We therefore regard the MP result as converged up to the dashed vertical line, obtained by requiring that none of the three truncation parameters $\mathcal{N}_{\max}$, $w$ and $\varepsilon_{\rm rel}$ changes $C(t)$ beyond a fixed tolerance, see the Supplemental Material \cite{supplement}. It is over this interval that the MP points appear in Fig.~\ref{fig:corrF}. There the two methods agree, confirming the recursion result. Going further in time is a question of memory rather than of principle: the largest runs already retain $5\cdot10^{8}$ strings, and this practical ceiling is typically reached before the decay of $C(t)$ is complete. Our numerical experiments confirm that MP is more efficient at small interaction strengths \cite{facelli2026fast,d2026majorana,Kuo_2024_Energy} and becomes increasingly demanding as $V$ grows \cite{facelli2026fast,alam2025fermionic}, a limitation absent in the recursion method. Another advantage of our implementation of the recursion method is that it works directly in the thermodynamic limit, whereas MP operates on finite lattices.

\bibliographystyle{unsrt}
\bibliography{ref}

\ifarXiv
    \foreach \x in {1,...,\numbersupplementpages}
    {
        \clearpage
        \includepdf[pages={\x}]{\supplementfilename}
    }
\fi

\end{document}